\documentclass[11pt]{article}
\usepackage{fullpage,epsf,amssymb,amsthm,amsfonts,amsmath,latexsym, graphicx,array,extarrows,mathtools,mathstyle,mathrsfs,slashed}
\usepackage[normalem]{ulem}

\newcommand{\overbar}[1]{\mkern 1.5mu\overline{\mkern-1.5mu#1\mkern-1.5mu}\mkern 1.5mu}

\let\oldsqrt\sqrt
\def\sqrt{\mathpalette\DHLhksqrt}
\def\DHLhksqrt#1#2{%
\setbox0=\hbox{$#1\oldsqrt{#2\,}$}\dimen0=\ht0
\advance\dimen0-0.2\ht0
\setbox2=\hbox{\vrule height\ht0 depth -\dimen0}
{\box0\lower0.4pt\box2}}

\numberwithin{equation}{section}

\title{\bf{Dyson-Schwinger equation constraints on the gluon propagator in BRST quantised QCD}} 
\author{\large{Peter Lowdon} \\
\ \\
\textit{\small{SLAC National Accelerator Laboratory, 2575 Sand Hill Rd, Menlo Park, CA 94025, USA}} \\
\textit{\small{E-mail: lowdon@slac.stanford.edu}}}
\date{}
\begin{document}
\begin{flushright}  SLAC-PUB-17219  \end{flushright}
\vspace{15mm} 
{\let\newpage\relax\maketitle}
\setcounter{page}{1}
\pagestyle{plain}

\abstract 

\noindent 
The gluon propagator plays a central role in determining the dynamics of QCD. In this work we demonstrate for BRST quantised QCD that the Dyson-Schwinger equation imposes significant analytic constraints on the structure of this propagator. In particular, we find that these constraints control the appearance of massless components in the gluon spectral density.

\newpage

\section{Introduction  \label{intro}}

Understanding the nature of confinement in QCD is crucial for explaining why quarks and gluons are absent from the physical spectrum of the theory~\cite{Alkofer_Greensite07}. Although there remains much debate surrounding the precise confinement mechanism, it has been understood for many years that the non-perturbative structure of the gluon propagator plays an important role~\cite{Mandula99}. An issue that has received significant focus in the literature is what happens to the propagator in the low momentum \textit{infrared} regime. Motivated by the issues surrounding gauge fixing, Gribov~\cite{Gribov78} and Zwanziger~\cite{Zwanziger89} proposed a form for the gluon propagator that vanishes in the limit $p^{2} \rightarrow 0$. Similar forms have also been proposed which suggest that the gluon propagator has an effective mass~\cite{Mandula_Ogilvie87}. In order to test both these and other hypotheses, a mixture of non-perturbative numerical and analytic techniques are often employed. In particular, the computation of the gluon propagator using lattice QCD and the Dyson-Schwinger equations remains a very active area of research~\cite{Alkofer_vonSmekal01,Alkofer_Detmold_Fischer_Maris04,Cucchieri_Mendes_Taurines05,Cucchieri_Mendes08,Strauss_Fischer_Kellermann12,Oliveria_Silva12, Dudal_Oliveira_Silva14}. Besides confinement, determining the structure of the gluon propagator is also important for describing other non-perturbative phenomena like the dynamics of quark-gluon plasma~\cite{Haas_Fister_Pawlowski14}, a topic which is currently the focus of significant theoretical and experimental interest at facilities such as ALICE (CERN) and RHIC (Brookhaven).  \\

\noindent
Many of the approaches to analysing the structure of the gluon propagator involve using the Becchi-Rouet-Stora-Tyutin (BRST) quantisation of QCD to work in specific Lorentz covariant gauges. BRST quantisation involves the introduction of additional auxiliary gauge-fixing and ghost degrees of freedom in such a way that the equations of motion are no longer gauge invariant, but remain invariant under a residual BRST symmetry. The physical states are then defined to be those that are annihilated by the conserved charge $Q_{B}$ associated with this symmetry~\cite{Nakanishi_Ojima90}. A key feature of BRST quantised QCD is that the space of states no longer possesses a positive-definite inner product, and hence negative norm states are permitted. This has the important implication that the momentum space correlation functions are no longer guaranteed to be non-negative~\cite{Bogolubov_Logunov_Oksak90}. Non-negativity violations of the gluon propagator are of particular relevance since this characteristic is often attributed to the absence of gluons from the physical spectrum~\cite{Alkofer_vonSmekal01,Oehme_Zimmermann80_1,Oehme_Zimmermann80_2,Cornwall13}, and recent numerical studies appear to indicate that these violations do indeed occur~\cite{Alkofer_Detmold_Fischer_Maris04,Cucchieri_Mendes_Taurines05,Strauss_Fischer_Kellermann12}. Although significant progress has been made in determining the structure of the BRST quantised gluon propagator, its behaviour remains far from understood. Part of the difficulty is that most of this progress has relied on functional techniques such as lattice QCD and the solution of the Dyson-Schwinger equations, both of which have significant uncertainties. A particularly prominant source of uncertainy concerns the non-perturbative definition of BRST symmetry, and whether this quantisation of QCD can in fact be implemented in spite of the Gribov problem~\cite{Gribov78,Zwanziger89}. \\

\noindent
In Ref.~\cite{Lowdon17_1} a more formal analytic approach was developed in order to determine the most general non-perturbative features of vector boson propagators. This approach involves the application of a rigorous quantum field theory framework, the construction of which is based on a series of physically motivated axioms~\cite{Nakanishi_Ojima90,Bogolubov_Logunov_Oksak90,Streater_Wightman64,Haag96,Strocchi13}. The advantage of this approach is that the axioms are assumed to hold independently of the coupling regime, and this allows genuine non-perturbative features to be derived in a purely analytic manner\footnote{Analytic approaches to constraining the non-perturbative structure of propagators have been pursued before, but have often relied on additional input such as the operator product expansion~\cite{Lowdon15_1}.}. For example, since BRST quantised QCD involves a space of states with an indefinite inner product, this opens up the possiblity that the gluon propagator contains singular terms involving derivatives of $\delta(p)$~\cite{Lowdon17_1}, a feature which is indicative of confinement~\cite{Strocchi76,Strocchi78,Lowdon16}. Nevertheless, it remains an open question as to whether the solutions of the gluon propagator derived using functional methods are actually sensitive to this type of singular behaviour. In this paper we adopt an axiomatic framework in order to provide a complimentary probe of the BRST quantised gluon propagator. Instead of solving the Dyson-Schwinger equation explicitly, we use this equation to derive analytic constraints on the form of this propagator.

\section{The gluon propagator in QCD}
\label{QCD_prop}

Before exploring the constraints that the Dyson-Schwinger equation imposes on the structure of the BRST quantised gluon propagator, it is important to first outline the dynamical characteristics of this theory, and the explicit form of the Dyson-Schwinger equation itself. The equations of motion of BRST quantised QCD are defined by
\begin{align}
&(D^{\nu}F_{\nu\mu})^{a} +\partial_{\mu}\Lambda^{a} = gj_{\mu}^{a} -igf^{abc}\partial_{\mu}\overbar{C}^{b}C^{c}, \hspace{3mm} \partial^{\mu}A_{\mu}^{a} = \xi\Lambda^{a}, \label{eom1} \\
&\partial^{\nu}(D_{\nu}C)^{a}=0, \hspace{5mm} (D^{\nu}\partial_{\nu}\overbar{C})^{a}=0, \label{eom2}
\end{align}
where $C^{a}$ and $\overbar{C}^{a}$ are the ghost and anti-ghost fields, $\Lambda^{a}$ is an auxiliary field, and $\xi$ is the renormalised gauge fixing parameter. It follows from Eq.~(\ref{eom1}) that the renormalised gluon field satisfies
\begin{align}
\left[ \partial^{2}g_{\mu}^{\ \alpha} - \left(1 - \frac{1}{\xi_{0}} \right)\partial_{\mu}\partial^{\alpha}  \right]A_{\alpha}^{a} = \mathcal{J}_{\mu}^{a}, 
\label{EOM_A}
\end{align}  
where $\xi_{0}$ is the bare gauge fixing parameter and $\mathcal{J}_{\mu}^{a}$ has the form
\begin{align}
\mathcal{J}_{\mu}^{a} =   gj_{\mu}^{a} -igf^{abc}\partial_{\mu}\overbar{C}^{b}C^{c} + (Z_{3}^{-1}-1)\partial_{\mu}\Lambda^{a} - igf^{abc}A^{b \nu}F_{\nu\mu}^{c} - gf^{abc}\partial^{\nu}(A_{\nu}^{b}A_{\mu}^{c}), 
\end{align}
with $Z_{3}$ the gluon field renormalisation constant and $j_{\mu}^{a}$ the matter current. Furthermore, one assumes that the renormalised fields satisfy the following equal-time commutation relations:
\begin{align}
&\left[\Lambda^{a}(x),\Lambda^{b}(y)\right]_{x_{0}=y_{0}} = 0, \hspace{5mm} \left[\Lambda^{a}(x),A_{\nu}^{b}(y)\right]_{x_{0}=y_{0}} = i\delta^{ab}g_{0\nu}\delta(\mathbf{x}-\mathbf{y}),  \label{etcr1} \\
&\left[A_{\mu}^{a}(x),A_{\nu}^{b}(y)\right]_{x_{0}=y_{0}} = 0, \hspace{5mm} \left[F_{0i}^{a}(x),A_{\nu}^{b}(y)\right]_{x_{0}=y_{0}} =  i\delta^{ab}g_{i\nu}Z_{3}^{-1}\delta(\mathbf{x}-\mathbf{y}). \label{etcr2}
\end{align}   
Since the gluon propagator is defined by
\begin{align}
\langle 0|T\{ A_{\mu}^{a}(x)A_{\nu}^{b}(y)\}|0\rangle  = \theta(x^{0}-y^{0}) \langle 0| A_{\mu}^{a}(x)A_{\nu}^{b}(y)|0\rangle + \theta(y^{0}-x^{0})\langle 0| A_{\nu}^{b}(y)A_{\mu}^{a}(x)|0\rangle,
\label{t_ordered_expl}
\end{align}
one can directly apply the dynamical conditions in Eqs.~(\ref{EOM_A}),~(\ref{etcr1}) and~(\ref{etcr2}) to this definition, and in doing so this implies the Dyson-Schwinger equation
\begin{align}
\left[ \partial^{2}g_{\mu}^{\ \alpha} - \left(1 - \frac{1}{\xi_{0}} \right)\partial_{\mu}\partial^{\alpha}  \right]\langle 0|T\{ A_{\alpha}^{a}(x)A_{\nu}^{b}(y)\}|0\rangle = i\delta^{ab} g_{\mu  \nu}Z_{3}^{-1} \delta(x-y) + \langle 0|T\{ \mathcal{J}_{\mu}^{a}(x)A_{\nu}^{b}(y)\}|0\rangle,
\label{SDE_x}
\end{align}
which in momentum space has the form
\begin{align}
-\left[ p^{2}g_{\mu}^{\ \alpha} - \left(1 - \frac{1}{\xi_{0}} \right)p_{\mu}p^{\alpha}  \right]\widehat{D}_{\alpha\nu}^{ab\, F}(p) = i\delta^{ab} g_{\mu  \nu}Z_{3}^{-1} + \widehat{J}_{\mu\nu}^{ab}(p),
\label{SDE_p}
\end{align}
where $\widehat{J}_{\mu\nu}^{ab}(p):= \mathcal{F}\left[\langle 0|T\{ \mathcal{J}_{\mu}^{a}(x)A_{\nu}^{b}(y)\}|0\rangle  \right]$. In what follows we will demonstrate that Eq.~(\ref{SDE_p}) imposes non-trivial analytic constraints on the structure of the gluon propagator. \\

\noindent
In order to explicitly understand the constraints imposed on the gluon propagator $\widehat{D}_{\mu\nu}^{ab\, F}(p)$ by Eq.~(\ref{SDE_p}), one must consider the spectral representation of both $\widehat{D}_{\mu\nu}^{ab\, F}(p)$ and the current propagator $\widehat{J}_{\mu\nu}^{ab}(p)$ involving the non-conserved current $\mathcal{J}_{\mu}^{a}$. In Ref.~\cite{Lowdon17_1} it was shown from Eqs.~(\ref{eom1}),~(\ref{etcr1}) and~(\ref{etcr2}) that the momentum space gluon propagator has the general form
\begin{align}
\widehat{D}_{\mu\nu}^{ab\, F}(p) &=   i\int_{0}^{\infty} \frac{ds}{2\pi} \, \frac{\left[ g_{\mu\nu}\rho_{1}^{ab}(s) + p_{\mu}p_{\nu}\rho_{2}^{ab}(s) \right]}{p^{2}-s +i\epsilon} -i \,g_{\mu 0}g_{\nu 0} \int_{0}^{\infty} \frac{ds}{2\pi} \, \rho_{2}^{ab}(s) \nonumber \\
 & \hspace{10mm} +\sum_{n=0}^{N} \left[ c_{n}^{ab} \, g_{\mu\nu} (\partial^{2})^{n} + d_{n}^{ab} \partial_{\mu}\partial_{\nu}(\partial^{2})^{n-1}\right]\delta(p),
\label{general_propagator_QCD_mom}
\end{align}
where $c_{n}^{ab}$ and $d_{n}^{ab}$ are complex coefficients which are linearly related\footnote{For $n=0$, $c_{n}^{ab}$ is unconstrained but $d_{n}^{ab}$ vanishes~\cite{Lowdon17_1}. As previously discussed, the possibility of non-vanishing terms involving derivatives of $\delta(p)$ arises because the BRST space of states has an indefinite inner product.} for $n \geq 1$, and the spectral densities $\rho_{1}^{ab}(s)$ and $\rho_{2}^{ab}(s)$ satisfy the following conditions~\cite{Lowdon17_1}
\begin{align}
\rho_{1}^{ab}(s) + s \rho_{2}^{ab}(s) = -2\pi\xi\delta^{ab}\delta(s), \hspace{2mm} \int_{0}^{\infty} ds  \, \rho_{1}^{ab}(s)  = -2\pi\delta^{ab} Z_{3}^{-1}, \hspace{2mm} \int_{0}^{\infty} ds  \, \rho_{2}^{ab}(s)  = 0.
\label{spectr_rel_gluon}
\end{align} 
The conditions in Eq.~(\ref{spectr_rel_gluon}) demonstrate that the gluon propagator contains only one independent spectral density\footnote{Subtleties can arise if one attempts to express the gluon propagator exclusively in terms of $\rho_{1}^{ab}(s)$~\cite{Lowdon17_1}, and this is why we will keep both spectral densities explicit in the proceeding analysis.}, and that the non-covariant term, which follows from the definition of the time-ordered product in Eq.~(\ref{t_ordered_expl}), actually vanishes due to the sum rule for $\rho_{2}^{ab}(s)$.  \\

\noindent
Now one can consider the structure of the propagator $\widehat{J}_{\mu\nu}^{ab}(p)$. The first constraint on this propagator arises from the fact that one can write the equations of motion for the gluon field as
\begin{align}
\partial^{\nu}F_{\nu\mu}^{a} = gJ_{\mu}^{a} + \left\{ Q_{B},(D_{\mu}\overbar{C})^{a} \right\},
\end{align}
where $\partial^{\mu}J_{\mu}^{a}=0$, and $Q_{B}$ is the BRST operator~\cite{Nakanishi_Ojima90}. By combining this equation with Eq.~(\ref{eom1}), the divergence of the current $\mathcal{J}_{\mu}^{a}$ can be written
\begin{align}
\partial^{\mu}\mathcal{J}_{\mu}^{a} =  \left\{Q_{B},Z_{3}^{-1}\partial^{2}\overbar{C}^{a} +(\partial^{\mu}D_{\mu}\overbar{C})^{a}\right\}.
\end{align} 
Using Eq.~(\ref{eom2}) together with the fact that $Q_{B}|0\rangle=0$, it then follows that the correlator $\langle 0| \mathcal{J}_{\mu}^{a}(x)A_{\nu}^{b}(y)|0\rangle$ satisfies the condition
\begin{align}
\partial^{\mu}_{x}\partial^{\nu}_{y}\langle 0| \mathcal{J}_{\mu}^{a}(x)A_{\nu}^{b}(y)|0\rangle = 0.
\label{vanish}
\end{align}
Using an analogous analysis as in the case of the gluon propagator~\cite{Lowdon17_1}, this condition implies that $\widehat{J}_{\mu\nu}^{ab}(p)$ has the same overall structural form     
\begin{align}
\widehat{J}_{\mu\nu}^{ab}(p) &=   i\int_{0}^{\infty} \frac{ds}{2\pi} \, \frac{\left[ g_{\mu\nu}\widetilde{\rho}_{1}^{ab}(s) + p_{\mu}p_{\nu}\widetilde{\rho}_{2}^{ab}(s) \right]}{p^{2}-s +i\epsilon}  +\sum_{n=0}^{\widetilde{N}} \left[ C_{n}^{ab} \, g_{\mu\nu} (\partial^{2})^{n} + D_{n}^{ab} \partial_{\mu}\partial_{\nu}(\partial^{2})^{n-1}\right]\delta(p),
\label{J_propagator_QCD_mom}
\end{align}
where $C_{n}^{ab}$ and $D_{n}^{ab}$ are complex parameters which are related in the same manner as $c_{n}^{ab}$ and $d_{n}^{ab}$ in Eq.~(\ref{general_propagator_QCD_mom}). Moreover, Eq.~(\ref{vanish}) implies that the spectral densities of this correlator are also not independent, and are in fact related as follows
\begin{align}
&\widetilde{\rho}_{1}^{ab}(s) + s \widetilde{\rho}_{2}^{ab}(s) = \widetilde{C}^{ab}\delta(s), \label{mix_rel}
\end{align}
where $\widetilde{C}^{ab}$ is a constant coefficient. In order to determine $\widetilde{C}^{ab}$, one can consider the contracted propagator expression $p^{\mu}p^{\nu}\widehat{J}_{\mu\nu}^{ab}(p)$, which due to Eqs.~(\ref{J_propagator_QCD_mom}) and~(\ref{mix_rel}) can be written
\begin{align}
p^{\mu}p^{\nu}\widehat{J}_{\mu\nu}^{ab}(p) = \frac{i}{2\pi}p^{2}\int_{0}^{\infty} ds \, \widetilde{\rho}_{2}^{ab}(s) + \frac{i}{2\pi}\widetilde{C}^{ab}.
\label{contr}
\end{align}     
Since $\widehat{J}_{\mu\nu}^{ab}(p)$ is defined by Eq.~(\ref{SDE_p}), contracting this equation with $p^{\mu}p^{\nu}$ gives an explicit expression for $p^{\mu}p^{\nu}\widehat{J}_{\mu\nu}^{ab}(p)$. In doing so, it follows from the Slavnov-Taylor identity\footnote{In this notation, the Slavnov-Taylor identity has the form $p^{\mu}p^{\nu}\widehat{D}_{\mu\nu}^{ab\, F}(p) = -i\xi\delta^{ab}$.} that 
\begin{align}
 p^{\mu}p^{\nu}\widehat{J}_{\mu\nu}^{ab}(p) =0, 
\end{align}
which in comparison with Eq.~(\ref{contr}) therefore implies the spectral density constraints
\begin{align}
&\widetilde{\rho}_{1}^{ab}(s) + s \widetilde{\rho}_{2}^{ab}(s) = 0 \hspace{5mm} (\widetilde{C}^{ab}=0), \label{constr_rho3_1} \\
& \hspace{10mm} \int_{0}^{\infty}ds \, \widetilde{\rho}_{2}^{ab}(s) = 0.
\label{constr_rho3_2}
\end{align}
 \ \\

\noindent
Having derived the spectral structure of both the gluon and current propagators, one can now determine the explicit constraints imposed by Eq.~(\ref{SDE_p}). Inserting Eqs.~(\ref{general_propagator_QCD_mom}) and~(\ref{J_propagator_QCD_mom}) into Eq.~(\ref{SDE_p}), and separately equating\footnote{Since the terms involving deriviatives of $\delta(p)$ have support only at $p=0$, whereas the other terms are defined to have support outside $p=0$ (in the closed forward light cone)~\cite{Bogolubov_Logunov_Oksak90}, this justifies why these terms can be separately equated.} the purely singular terms involving derivatives of $\delta(p)$, one obtains
\begin{align}
&\left[ -p^{2}g_{\mu}^{\ \alpha} + \left(1 - \frac{1}{\xi_{0}} \right)p_{\mu}p^{\alpha}  \right]\left[\sum_{n=0}^{N} \left[ c_{n}^{ab} \, g_{\alpha\nu} (\partial^{2})^{n} + d_{n}^{ab} \partial_{\alpha}\partial_{\nu}(\partial^{2})^{n-1}\right]\delta(p) \right] \nonumber \\
& \hspace{50mm} = \sum_{n=0}^{\widetilde{N}} \left[ C_{n}^{ab} \, g_{\mu\nu} (\partial^{2})^{n} + D_{n}^{ab} \partial_{\mu}\partial_{\nu}(\partial^{2})^{n-1}\right]\delta(p), \label{constr_2} \\
&\left[ -p^{2}g_{\mu}^{\ \alpha} + \left(1 - \frac{1}{\xi_{0}} \right)p_{\mu}p^{\alpha}  \right]\left[i\int_{0}^{\infty} \frac{ds}{2\pi} \, \frac{\left[ g_{\alpha\nu}\rho_{1}^{ab}(s) + p_{\alpha}p_{\nu}\rho_{2}^{ab}(s) \right]}{p^{2}-s +i\epsilon} \right]  \nonumber \\
& \hspace{50mm} = i\delta^{ab} g_{\mu  \nu}Z_{3}^{-1} + \left[ i\int_{0}^{\infty} \frac{ds}{2\pi} \, \frac{\left[ g_{\mu\nu}\widetilde{\rho}_{1}^{ab}(s) + p_{\mu}p_{\nu}\widetilde{\rho}_{2}^{ab}(s) \right]}{p^{2}-s +i\epsilon} \right]. \label{constr_1}
\end{align}
Expanding out the left-hand-side of Eq.~(\ref{constr_2}) it follows that the coefficients $c_{n}^{ab}$ and $d_{n}^{ab}$ are directly related to $C_{n}^{ab}$ and $D_{n}^{ab}$. In particular, one has the relation
\begin{align}
c_{n+1}^{ab} = -\frac{(2n+5)}{4(2n+3)(n+1)(n+3)}C_{n}^{ab}, \hspace{5mm} n \geq 0  \label{c_constr_rel} 
\end{align} 
Since both $c_{n}^{ab}$, $d_{n}^{ab}$, and $C_{n}^{ab}$, $D_{n}^{ab}$ are separately linearly related, Eq.~(\ref{c_constr_rel}) implies that all of these parameters must be linearly related to one another. The significance of these relations is that they demonstrate that the coefficients of terms involving derivatives of $\delta(p)$ in the gluon propagator ($c_{n}^{ab}$ and $d_{n}^{ab}$ for $n \geq 1$) are proportional to the coefficients of $\delta(p)$ and derivatives of $\delta(p)$ in $\widehat{J}_{\mu\nu}^{ab}(p)$. In particular, for $n=0$ Eq.~(\ref{c_constr_rel}) implies that if $\widehat{J}_{\mu\nu}^{ab}(p)$ has a non-vanishing $\delta(p)$ term, this is sufficient to prove that the gluon propagator contains a $\partial^{2}\delta(p)$ component. This characteristic is particularly relevant in the context of confinement, since the appearance of singular terms involving non-vanishing derivatives of $\delta(p)$ is related to the violation of the cluster decomposition property~\cite{Strocchi76, Strocchi78, Lowdon16, Nakanishi_Ojima90,Roberts_Williams_Krein91}. Eq.~(\ref{c_constr_rel}) therefore demonstrates that the singular structure of the interaction current propagator $\widehat{J}_{\mu\nu}^{ab}(p)$ plays an important role in understanding this phenomenon. \\

\noindent
In order to derive the constraints imposed by Eq.~(\ref{constr_1}), one must expand this equation and then separately equate the terms on both sides which depend on $g_{\mu\nu}$ and $p_{\mu}p_{\mu}$. In doing so, this implies the relations
\begin{align}
&-p^{2}\int_{0}^{\infty} \frac{ds}{2\pi} \, \frac{\rho_{1}^{ab}(s)}{p^{2}-s +i\epsilon} = \delta^{ab} Z_{3}^{-1} + \int_{0}^{\infty} \frac{ds}{2\pi} \, \frac{\widetilde{\rho}_{1}^{ab}(s)}{p^{2}-s +i\epsilon}, \label{constr_1_1} \\
&\int_{0}^{\infty} \frac{ds}{2\pi} \, \frac{\left(1 - \frac{1}{\xi_{0}} \right)\rho_{1}^{ab}(s)  - \frac{1}{\xi_{0}}p^{2}\rho_{2}^{ab}(s)     }{p^{2}-s +i\epsilon} =  \int_{0}^{\infty} \frac{ds}{2\pi} \, \frac{\widetilde{\rho}_{2}^{ab}(s)}{p^{2}-s +i\epsilon}. \label{constr_1_2}
\end{align}
Using the constraints in Eq.~(\ref{spectr_rel_gluon}), it follows from Eq.~(\ref{constr_1_1}) that $\rho_{1}^{ab}(s)$ satisfies the equality
\begin{align}
s\rho_{1}^{ab}(s) + \widetilde{\rho}_{1}^{ab}(s) = 0,
\label{rho1_c}
\end{align}
which in combination with Eq.~(\ref{constr_rho3_1}) implies
\begin{align}
s\left[\rho_{1}^{ab}(s) - \widetilde{\rho}_{2}^{ab}(s)\right] = 0.
\label{cond_zero}
\end{align}
In order to solve this equation it is important to recognise that because spectral densities are distributions, not functions, the solution is not necessarily continuous\footnote{See Ref.~\cite{Lowdon17_1} for a general discussion of this issue.}. In fact, the general solution of Eq.~(\ref{cond_zero}) has the form: $\rho_{1}^{ab}(s) - \widetilde{\rho}_{2}^{ab}(s) = A^{ab}\delta(s)$, where $A^{ab}$ is a constant coefficient~\cite{Bogolubov_Logunov_Oksak90}. By applying the integral constraints in Eqs.~(\ref{spectr_rel_gluon}) and~(\ref{constr_rho3_2}) this fixes the coefficient to: $A^{ab}= -2\pi \delta^{ab}Z_{3}^{-1}$, and hence
\begin{align}
\rho_{1}^{ab}(s) = -2\pi \delta^{ab}Z_{3}^{-1}\delta(s) + \widetilde{\rho}_{2}^{ab}(s).  \label{constr_rho1} 
\end{align}  
Applying an analogous approach to Eq.~(\ref{constr_1_2}) subsequently leads to the following constraint 
\begin{align}
s\rho_{2}^{ab}(s) = 2\pi \delta^{ab}\left( Z_{3}^{-1} -\xi \right) \delta(s) - \widetilde{\rho}_{2}^{ab}(s). \label{constr_rho2}
\end{align}
\ \\
\noindent
In general, Eqs.~(\ref{constr_rho1}) and~(\ref{constr_rho2}) demonstrate that the behaviour of the gluon spectral densities is completely determined by the spectral densities of the current propagator $\widehat{J}_{\mu\nu}^{ab}(p)$. Moreover, Eq.~(\ref{constr_rho1}) implies that $\rho_{1}^{ab}(s)$ contains an explicit massless contribution, which has an overall $Z_{3}^{-1}$ coefficient. Since $Z_{3}^{-1}$ is expected to vanish in Landau gauge~\cite{Alkofer_vonSmekal01}, massless gluons must therefore necessarily be absent from the spectrum in this gauge. However, because $Z_{3}^{-1}$ is gauge dependent, the absence of a massless gluon component is not necessarily guaranteed in other gauges\footnote{Performing the same analytic procedure for the photon propagator would also result in a massless spectral density component with a $Z_{3}^{-1}$ prefactor, where now $Z_{3}$ is the photon field renormalisation constant, which is gauge invariant.}. In the literature~\cite{Alkofer_Detmold_Fischer_Maris04,Cucchieri_Mendes_Taurines05, Strauss_Fischer_Kellermann12,Dudal_Oliveira_Silva14,Cornwall13} it is often argued that the violation of non-negativity of $\rho_{1}^{ab}(s)$ in Landau gauge as a result of the sum rule\footnote{This sum rule is often referred to as the Oehme-Zimmermann superconvergence relation~\cite{Oehme_Zimmermann80_1,Oehme_Zimmermann80_2}.}: $\int ds \, \rho_{1}^{ab}(s)=0$ is the reason why gluons do not appear in the spectrum. However, from the structure of Eq.~(\ref{constr_rho1}) it is apparent that (continuous) non-negativity violations can only arise from the component $\widetilde{\rho}_{2}^{ab}(s)$, which has vanishing integral [Eq.~(\ref{constr_rho3_2})]. Since the analogous component $\widetilde{\rho}_{2}(s)$ of the photon spectral density in QED turns out to also have vanishing integral, this implies that potential non-negativity violations are not QCD specific, and casts doubt on the hypothesis that these violations in Landau gauge are the reason why gluons are absent from the spectrum.

\section{Conclusions}

In this work we have demonstrated for the first time that the Dyson-Schwinger equation imposes non-trivial analytic constraints on the structure of the gluon propagator in BRST quantised QCD. These constraints imply that the gluon spectral density explicitly contains a massless component, but that the coefficient of this component is gauge-dependent. As well as the purely theoretical relevance of this result, these constraints could also provide important input for improving existing parametrisations of the gluon propagator.

\section*{Acknowledgements}
I thank Reinhard Alkofer for useful discussions and input. This work was supported by the Swiss National Science Foundation under contract P2ZHP2\_168622, and by the DOE under contract DE-AC02-76SF00515.

\renewcommand*{\cite}{\vspace*{-12mm}}

\end{document}